\documentclass[useAMS,usenatbib,referee]{mn2e}
\usepackage{graphicx}  \usepackage[dvips,letterpaper]{geometry}
\usepackage{color}

\def\msun{\ensuremath{M_{\odot}}}

\def\th{\tau_{H(4)}}
\def\simlt{\lower.5ex\hbox{$\; \buildrel < \over \sim \;$}}
\def\simgt{\lower.5ex\hbox{$\; \buildrel > \over \sim \;$}}
\def\taustd{\ensuremath{\tau_{\mathrm{std}}}}

\def\uc05{UC05}

\title[Recombination in supernovae]{
On the Hydrogen Recombination Time in Type II Supernova Atmospheres
}

\author[De, Baron, \& Hauschildt]{
Soma De$^{1}$, E.~Baron$^{1,2}$, P.~H.~Hauschildt$^{3}$\\
$^{1}$Homer  L. Dodge Dept.   of  Physics and Astronomy, University of Oklahoma,
Norman, OK 73019, USA\\ 
$^{2}$Computational Research Division, Lawrence Berkeley
        National Laboratory, MS 50F-1650, 1 Cyclotron Rd, Berkeley, CA
        94720 USA\\
$^{3}$Hamburger Sternwarte, Gojenbergsweg 112, 21029 Hamburg, Germany\\
}

\begin{document}

\pubyear{2009}

\maketitle

\label{firstpage}

\begin{abstract}
  NLTE radiative transfer calculations of differentially expanding
  supernovae atmospheres are computationally intensive and are almost
  universally performed in time-independent snapshot mode, where both
  the radiative transfer problem and the rate equations are solved
  assuming the steady-state approximation. The validity of the
  steady-state approximation in the rate equations has recently been
  questioned for Type II supernova (SN II) atmospheres after maximum
  light onto the plateau.  We calculate the effective recombination
  time of hydrogen in SN~II using our general purpose model atmosphere
  code \texttt{PHOENIX}.  While we find that the recombination time
  for the conditions of SNe~II at early times is increased over the
  classical value for the case of a simple hydrogen model atom with
  energy levels corresponding to just the first 2 principle quantum
  numbers, the classical value of the recombination time is recovered
  in the case of a multi-level hydrogen atom.  We also find that the
  recombination time at most optical depths is smaller in the case of
  a multi-level atom than for a simple two-level hydrogen atom. We
  find that time dependence in the rate equations is important in the
  early epochs of a supernova's lifetime. The changes due to the time
  dependent rate equation (at constant input luminosity) are
  manifested in physical parameters such as the level populations
  which directly affects the spectra. The $H_\alpha$ profile is
  affected by the time dependent rate equations at early times. At
  later times time dependence does not significantly modify the level
  populations  and therefore the $H_\alpha$ profile is
  roughly independent of whether the steady-state or time-dependent
  approach is used.
\end{abstract}
 

\section{Introduction}

Supernovae are one of the most widely researched objects in
astrophysics.  There has been much debate about the detailed physical
nature of these objects, driven by the discovery of the dark energy
using SNe Ia \citep{riess_scoop98,perletal99}. Supernovae add rich
variety to the metal abundances in galaxies and are important to star
formation theories. Supernova nucleosynthesis of heavy elements is
responsible for galactic chemical evolution.  Prior to the discovery
of dark energy, supernova research was spurred on by the discovery of
the peculiar Type II SN 1987A in the LMC \citep[see][and references
therein]{sn87arev89}.  Type II 
supernovae (SNe II) are classified by the presence of strong Balmer
lines in their spectra in contrast with SNe I which lack strong Balmer
lines.  SNe II are due to core collapse of massive stars.  SN~1987A
was extremely well observed and much work has been done to explain all
aspects of its spectra. The Balmer lines were not well reproduced at
early times (especially about 8 days after explosion).  At first
very large overabundances of the s-process elements Ba and Sc were
suggested \citep{williams87A87,williams87A87b,hof87A88,danziger87A88}.
Further work showed that the 
Ba~II and Sc~II lines could be  reproduced by assuming that the
s-process abundances were enhanced by a factor of 5 compared to the LMC
abundances \citep{mazz87A92}.  

The  hydrogen Balmer
  line problem in many SNe~II has not been accounted in a
  consistent manner.  \citet{dessart08} summarizing their work using
  time-independent rate equations 
   found that they were unable to reproduce the $H_\alpha$ line 
  after 4~days in SN 1987A, 40~days in SN~1999em, and 20~days in SN~1999br.
   \citet{Chugai}
  point out the lack of a consistent 
  physically motivated framework to explain the early epoch Balmer
  profile that matches with the observed line
  strength. Other authors followed different approaches to reproduce
  the observed line strengths with varying degrees of success
  \citep{h03a,schm90,HM98,dessart05b}.
\citet{Chugai} argued that the poor fit of the Balmer lines could be
overcome by including time-dependence into the hydrogen rate
equations.  They argued that even in the intermediate layers, the
recombination time increases due to Lyman alpha trapping or ionization
from the first excited state.  Consequently, the recombination time
becomes comparable to the age of the supernova, making the
steady-state approximation suspect.  We numerically estimated the
recombination time by directly calculating the recombination rate from
the continuum into the bound states of hydrogen using H in NLTE with
31 bound states and all other metals in LTE.  We use the general
purpose model atmosphere code \texttt{PHOENIX} developed by
\citet{hbjcam99}.  We specify the density structure $\rho \propto
r^{-7}$ and assume the ejecta mass to be 14.7~\msun\ as is appropriate
for SN~1987A \citep{nomoto1988}.  \texttt{PHOENIX} calculates the
level populations, temperature, and radiation field for successive
epochs with a time interval of approximately two days.  We generated
our models and spectra using both time-dependent and time-independent
rate equations.  We calculated our recombination time for each epoch
from the best fit spectra in each case.  In \S~2, we outline the
theoretical framework for the recombination time and motivation for
our work.  In \S~3, we discuss earlier work done to test the relevance
of time-dependence in type II supernova atmospheres. In \S~4, we
describe the basic framework for our code and in \S~5 we present our
approach to compute the recombination time. In \SS~6--8 we analyze our
results and address the question of 
the necessity of incorporating time dependence in the rate equations
due to the increased recombination time.

\section{Motivation}

The recombination time is given by \citep{osterbrock89}
\begin{equation}
 \tau_{rec} = \frac{1}{n_e \alpha_A} = \frac{3\times 10^{12}\,\mathrm{s}}{n_e}
 \label{simple_tau}
\end{equation}
where $n_e$ is the number density of electrons and $\alpha_A$ is the
recombination coefficient. The Case A recombination coefficient is
used for convenient comparison with \citet{Chugai} who also assume
Case A.  \citet{HS87} suggest that the Case B recombination
coefficient is not appropriate for cases with Lyman~$\alpha$ escape or
with the electron densities as high as $10^{8}$~g~cm$^{-3}$. Thus,
Case A is more relevant here and since \citet{Chugai} use the same
coefficient we will use it to facilitate comparison.

It has been recently argued by \citet{Chugai} (henceforth \uc05) that
time dependence in the rate equations in a normal SN IIP atmosphere
may be important due to a significant increase in the recombination
time. It is well-known that in the early universe Lyman-alpha trapping
and ionization from the second level increases the effective
recombination time in hydrogen \citep{Zeldovich,Peebles}. The net
recombination rate is then determined by transitions that are not as
highly resonant as L$\alpha$ such as many non-resonant processes
connecting the same parity states (for example the 2$\gamma$ process)
and the processes that are accompanied by the escape of the resonant
photon.  For a hydrogen model atom with $n=2$ (consisting of only
levels 1$s$, 2$s$, 2$p_{\frac{1}{2}}$ and 2$p_{\frac{3}{2}}$), the
ionization from the second level dominates the L$\alpha$ escape
probability or the 2$\gamma$ transition probability from the 2$s$ to
the 1$s$ level. This delays the effective recombination or lengthens
the recombination time. In the case of an atom with many excited
levels, recombination into the upper levels followed by fluorescence
can significantly alter the ionization fraction due to the well known
phenomena of ``photon suction'' \citep*{CRSMgI92}. The
  escape probability for the resonant Lyman lines 
  increases as the energy of the bound
  state increases (S.~De et al., in preparation). Thus, since these
  photons escape, the electrons end up in the ground state enhancing
  recombination. 
These two effects (in
multi-level atoms), could affect the recombination process
significantly, altering the physical conditions considered by \uc05,
so that there is once again efficient recombination and one should not
necessarily expect any significant increase in recombination time over
that given by Eqn~\ref{simple_tau}.  In the following sections we
briefly describe relevant earlier work and explain why it is important
to check the numerical value of the recombination time in the
system. To be able to correctly calculate the recombination rate for a
multilevel atom we use \citep{Peebles}
\begin{equation}
-\frac{d}{dt}\left ( \frac{n_{e}}{n} \right ) = \sum \frac{(R_{nl}-P_{nl})}{n}.
\end{equation}
$R_{nl}$ is the recombination rate into any bound level characterized
by the principle quantum number $n$ and angular quantum number $l$,
from the continuum and $P_{nl}$ is the photo-ionization rate from any
bound level into the continuum. This is similar to 
equation (23) of \citet{Peebles}. The difference is that instead of
using simplified assumptions, we calculate the net recombination rate
from the simultaneous solution of the rate equation and the radiative
transfer equation.  We then calculate the recombination time for a
multi-level atom model by using
\begin{equation}
\tau_{rec} = \frac{n_{e}}{\sum (R_{nl}-P_{nl})} 
\label{equation1}
\end{equation}
Equation~\ref{equation1} is an equivalent way of
  estimating the recombination time.  
  \citet{Chugai}  used 
\begin{equation}
\tau_{rec}=\frac{n_{e}}{\frac{dn_{e}}{dt}}
\label{eq:chug_rectime}
\end{equation}
to estimate the recombination time.
In  Eqs.~\ref{equation1} and \ref{eq:chug_rectime}, when
the Lagrangian derivative of 
the free electron density with respect to time goes to zero (in the case
of ionization freeze out), the recombination time goes to
infinity. This contradiction is only apparent, since this equation 
only computes the time-scale associated with the variable part of the
free electron density. When the free electron density is constant in
the Lagrangian frame, this equation implies that the time associated with
any variation in the free electron density is infinite or in other
words, it does not vary. 

Since we are
really considering just the recombination of hydrogen, in a solar
mixture an alternative would be to define
\begin{equation}
\tau_{rec}= \frac{n_{H^{+}}}{R_{nl}-P_{nl}} 
\label{equation2}
\end{equation}
where $n_{H^{+}}$ is the proton density in the system. The reason why the
electron recombination time scales with hydrogen ion density is
because we treat other elements except hydrogen in LTE or in other
words the other elements  do not have explicit time dependence in their
ionic concentrations.  Generally 
at the relevant optical depths, the ratio
$n_{H^{+}}/n_{e}$ is very close to 1.0, so there is almost no
difference between
Eqs.~\ref{equation1} and \ref{equation2}. We will therefore use
Equation~\ref{equation1}.

\section{ Earlier work}

There have been different numerical models that treat the problem of
recombination. This problem of recombination was  addressed by
\citet{Zeldovich} in the cosmological recombination epoch
scenario. As noted above, \citet{Chugai} have revived this  for the
case of SNe~II.
\citet{dessart08} calculated SNe~IIP spectra
obtained treating the rate equations including 
time-dependent effects. We try to carefully review the effect
of time-dependence in the results published by \citet{dessart08}.
The main idea is to check if the temporal nature of the rate
equations is significantly supported by \uc05's argument
on the recombination time.

\subsection{Utrobin and Chugai's calculation}

\uc05 used a two-level plus continuum approximation for the hydrogen
atom  to estimate the
recombination probability and then estimate the recombination
time. They argue that around $N_{e} > 10^{8}$ and neutral hydrogen
number density $N_{HI} > 10^{9}$,  two-photon process dominates over
other processes (collisional terms or the escape probability term) and
the recombination probability is computed to be $< 2.0 \times
10^{-3}$. The characteristic temperature was 5000~K, the
photo-ionization probability was assumed to be $\sim 10^{3}$~s$^{-1}$,
and the dilution factor was taken as $W= 0.1$.  Therefore the classical
recombination time $(\alpha N_{e})^{-1}$ gets modified to
$\frac{1}{\alpha N_{e} w_{21}}$ which is factor of 500 higher than the
classical recombination time.  For a typical type II
supernova with age $10^{6} -10^{7}$~s, the modified
recombination time becomes of the order $10^{7}$~s. To test
their numerical estimate and predictions about the importance of
incorporating time-dependence, \uc05 treated the radiation field in the
core-halo approximation and assumed the Sobolev approximation for line
formation. Assuming that the SN atmosphere is very opaque in the Lyman
continuum enabled them to fully determine the diffusive continuum in
that frequency band by hydrogen recombinations and free-free
emissions.

At lower frequencies, between the Balmer and the Lyman edges, there is
a large optical depth owing to  interaction with numerous
metal lines. Under these conditions \uc05 assumed that absorption 
was given by the constant absorption coefficient
from the solution of
\citet{Chandra}.

In the visual band where the optical depth of the atmosphere is quite
low, they adopted the free streaming approximation, which describes
the average intensity of the continuum in the atmosphere as 
proportional to the specific intensity of photospheric radiation. The
effective radius and temperature are given by their hydrodynamical
model which was created after SN 1987A. \uc05 incorporated gamma-ray
deposition and included the time-dependent term in their
rate equations for all species they considered.

They reproduced the H$_\alpha$ line that appears in the photospheric
epoch in the first month of 
SN 1987A. They were unable to reproduce the
additional blue peak that appears between day 20 and 29 which they
argued was due to the Bochum event \citep{dachs}. Also they were able
to reproduce the Ba II $\lambda 6142$  line in SN 1987A between days 14 and
19 with LMC barium abundances. They argued that time-dependent
hydrogen ionization provided higher electron densities in the
atmosphere and thus made recombination of Ba~III into Ba~II more
efficient.
\subsection{ Dessert and Hillier's work}

\citet{dessart08} incorporated  time dependent terms into the
statistical and radiation equilibrium calculations of the non-LTE line
blanketed radiative transfer code CMFGEN. They allowed full
interaction between the radiation field and level populations to study
the effect on the full spectrum.  \citet{dessart08}
discuss their findings on the ejecta properties and spectroscopic
signatures obtained from time dependent simulations. They neglected
time dependent and relativistic terms in the radiative transfer
equation.  However, they argued that inclusion of those terms would not
affect their results because the importance of the time dependent
terms arises primarily due to atomic physics and should not be
sensitive to radiative transfer effects.

They compare their results with a sample of observations. They
reported a strong and broad H$_\alpha$ line that closely matches the
observed profile for SN 1999em in the hydrogen recombination epoch
\emph{without} the inclusion of non-thermal ionization/excitation due
to gamma-ray deposition from the radioactive decay of $\
^{56}$Ni.  They
were also able to reproduce the H$_\alpha$, Na~I D and Ca II IR triplet
($\sim 8500$~\AA) lines more satisfactorily.

Figure 16 of \citet{dessart08} compares the observed spectra of SN
1999em for 48.7 days since explosion and the 
spectra calculated from both time dependent and independent models
for the rate 
equations.  One may notice a significant mismatch between the time
dependent model and observed spectra in certain regions where the time
independent models matches better.  Around 4000 \AA, the Fe~I, Fe~II
lines are not reproduced well \citep{Pastorello}, similarly 
the C~I and Ca~II line strengths around 8600~\AA\ are underestimated
using the time dependent model and the line strengths are better
reproduced by the time independent model. 
\citet{dessart08} find that time dependence is better for later
epochs, from their work on SN 1999em and SN 1999br,
about 30 days after explosion.  

\section{Description of \texttt{PHOENIX}}

\texttt{PHOENIX} is a model atmosphere computer code that has been
developed by Hauschildt, Baron and collaborators over the last two
decades. Due to the coupling between the level populations and the
radiation field, the radiation transport equation must be solved
simultaneously with the rate equations. \texttt{PHOENIX} includes a
large number of NLTE and LTE background spectral lines and solves the
radiative transfer equation with a full characteristics piecewise
parabolic method \citep{ok87,phhs392} and \emph{without} simple
approximations like the Sobolev approximation \citep{Mihalas}.
\citet{hbjcam99} describe the numerical algorithms used in
\texttt{PHOENIX} to solve the radiation transport equations, the
non-LTE rate equations and parallelization of the
code. \texttt{PHOENIX} uses the total bolometric luminosity in the
observer's frame, the density structure, and element abundances as
input parameters. The equation of radiative transfer in spherical
geometry, the rate equations, and the condition of radiative
equilibrium are then solved. This process is repeated until the
radiation field and the matter have converged to radiative equilibrium
in the Lagrangian frame.  In our model calculations
  we have not taken into account the ionization by  non thermal
  electrons. Also time dependence was incorporated only in the
  rate equations. The inclusion of time dependence in the thermal
  equilibrium equation is implicit. Since the time dependent rate
  equations affect the opacities, the solution of the radiative transfer
  equation is affected as well. The explicit implementation of time
  dependence in the radiative transfer equation is beyond the scope of
  this work. Our goal was to specifically examine the effects
  of time dependence on the hydrogen Balmer lines and therefore only hydrogen is treated in NLTE. 
  This both speeds
  up the calculations significantly and isolates the effects of time
  dependence on hydrogen.

\section{Recombination Time Calculations}

Our main motivation is  to explore the
recombination time in a SN II atmosphere. In order to obtain the
correct electron densities and level populations it is essential to
make sure that the rate equations are correct even if the
recombination time is actually longer or comparable to the age of the
supernova. We incorporated the time dependent term in the rate
equations and in our results we describe this as the time dependent
case. We have performed calculations assuming both the steady-state,
time-independent, (TI), approximation as well as the full time-dependent,
(TD), rate equations.
In
most of our models we used a hydrogen model atom consisting of 31
levels. To save computing time and limit the time-dependent effects
to the Balmer lines, we treated all
other elements in LTE for this work. We started with a known density profile
typical of a SN II with mass similar to the progenitor mass of
SN~1987A ($\sim 14.7\msun$) and luminosity ($1.015 \times$ $10^{43}$
ergs) . We built a radial velocity grid with 128 layers. 
The location of the radial points was found from the assumption of homology,
$r=vt$. \texttt{PHOENIX}  then  solves the
radiative transfer equation and the rate equations
simultaneously. This yields the  temperature, radiation field,
and other physical parameters such as pressure, optical depth, and
electron density for each radial grid or layer. We choose the best
model by tuning only the input luminosity and the input density profile by
comparing the match between the synthetic and the observed spectra of
SN~1987A.

We began  with a time-independent model at  Day 2 and generated
snapshots for both the time 
dependent and independent cases up to Day 20 with intervals of 2 days.
It is important for the time-dependent case to have the
correct level populations at the previous time and consequently it was
better to choose a smaller time interval (two days) to generate
snapshots of  the physical profile of the supernova. Our analysis was done
on a Lagrangian velocity grid. 
Right after the explosion, the optical depth is very high
in the innermost layers that are expanding with low
velocity. At very high optical depth, radiative diffusion is an
excellent approximation. Thus at high optical depths we replace the
innermost ejecta with an opaque core and the luminosity is given by
the expression for radiative diffusion at this boundary. 
Our initial grid has 
 has a velocity of $ \sim 3000 $~km~s$^{-1}$ and we gradually add
 deeper layers until at our final epoch of 
 20 days, the innermost layer has a  velocity of around
$\sim 1000$~km~s$^{-1}$. To add new layers in the inner core we merged
 the outer low density, high velocity layers keeping the
total number of layers constant. This way we preserved the grid size
but also made sure that we have the physically relevant velocity
range at all epochs of evolution.

In order to recover the results of \uc05 in a simple hydrogen model atom case 
and also to understand the difference between the simple and
multi-level model atom framework we repeated our calculations for a
4-level hydrogen atom model case with solar compositions.
In order to have the correct physical conditions, we held the
structure of the atmosphere (temperature and density) fixed from what
we had obtained with the time-dependent multi-level hydrogen atom case  for that epoch.
We then solved for the time-dependent level populations and the
associated radiation field.

\section{Results}

Our computations consist of 
three different systems at each epoch.
They are  the case when we have a
4-level ($n=2$) hydrogen model atom in NLTE (in absence of any other metals)
and time-dependent rate equations. We refer to this model as Case 1.
Cases 2 and 3 are full calculations with 31-level hydrogen model atoms,
solar compositions, and 
time-dependent and time-independent rate equations,
respectively. 

We measure the optical depth using the value $\tau_{\mathrm{std}}$
which is the optical depth in the continuum at 
5000~\AA.

\subsection{Ionization Fraction and Electron Density Profile}

\begin{figure*}
\centering
\includegraphics[width=0.65\textwidth,angle=90]{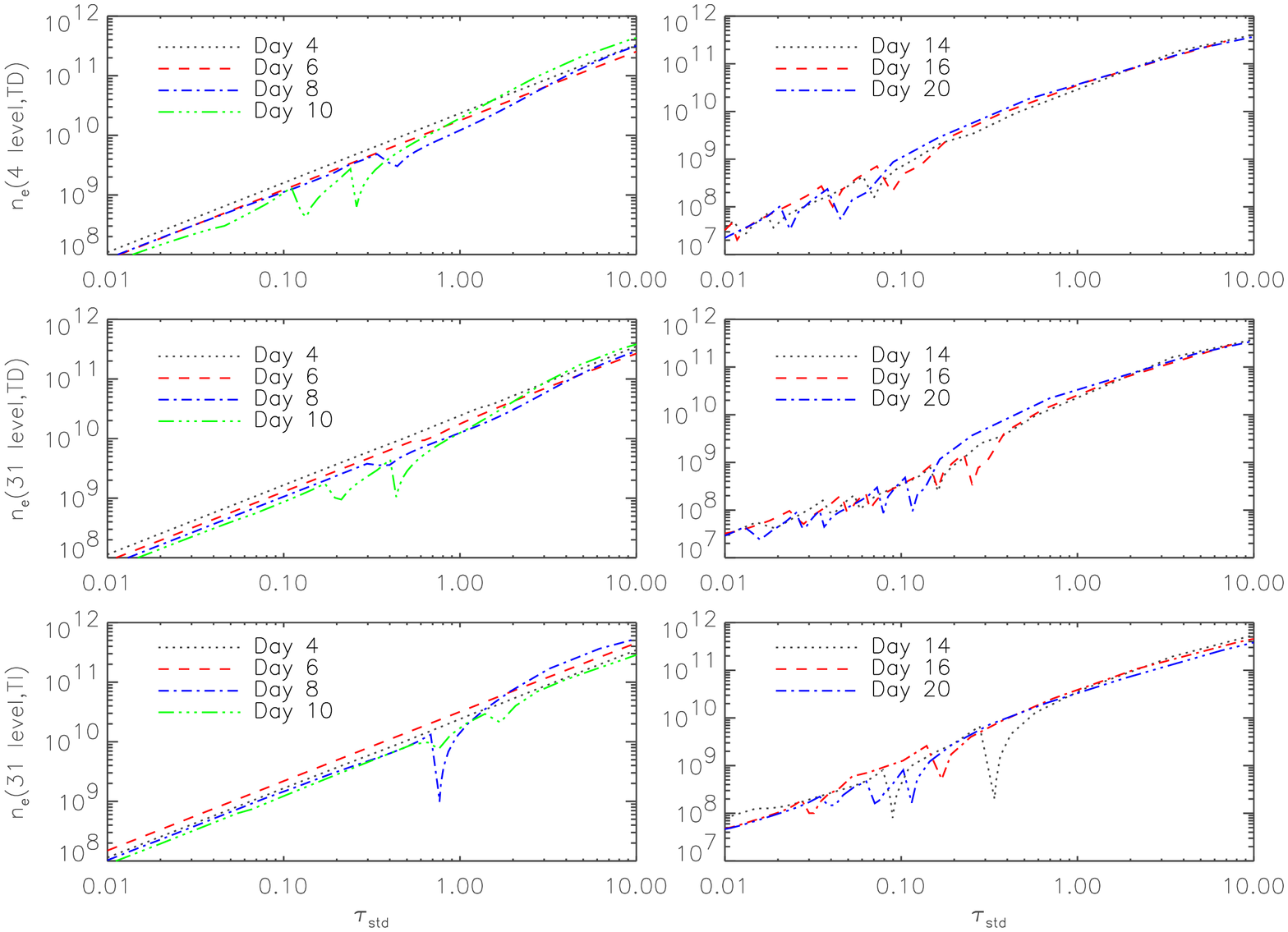}
\caption{Comparison between the free electron densities at different
  systems. The upper panel has 4-level hydrogen atom case. The middle
  panel has the multilevel time dependent case. The bottom panel has
  the multilevel time independent case. 
  \label{electron_density}}
\end{figure*}

The two most important physical parameters for the 
recombination time are the ionization fraction, $f_H$, and the free
electron density, $n_e$. In Figure~\ref{electron_density} we present the
electron density for the three different cases.

\begin{figure*}
\centering
\includegraphics[width=0.65\textwidth,angle=90]{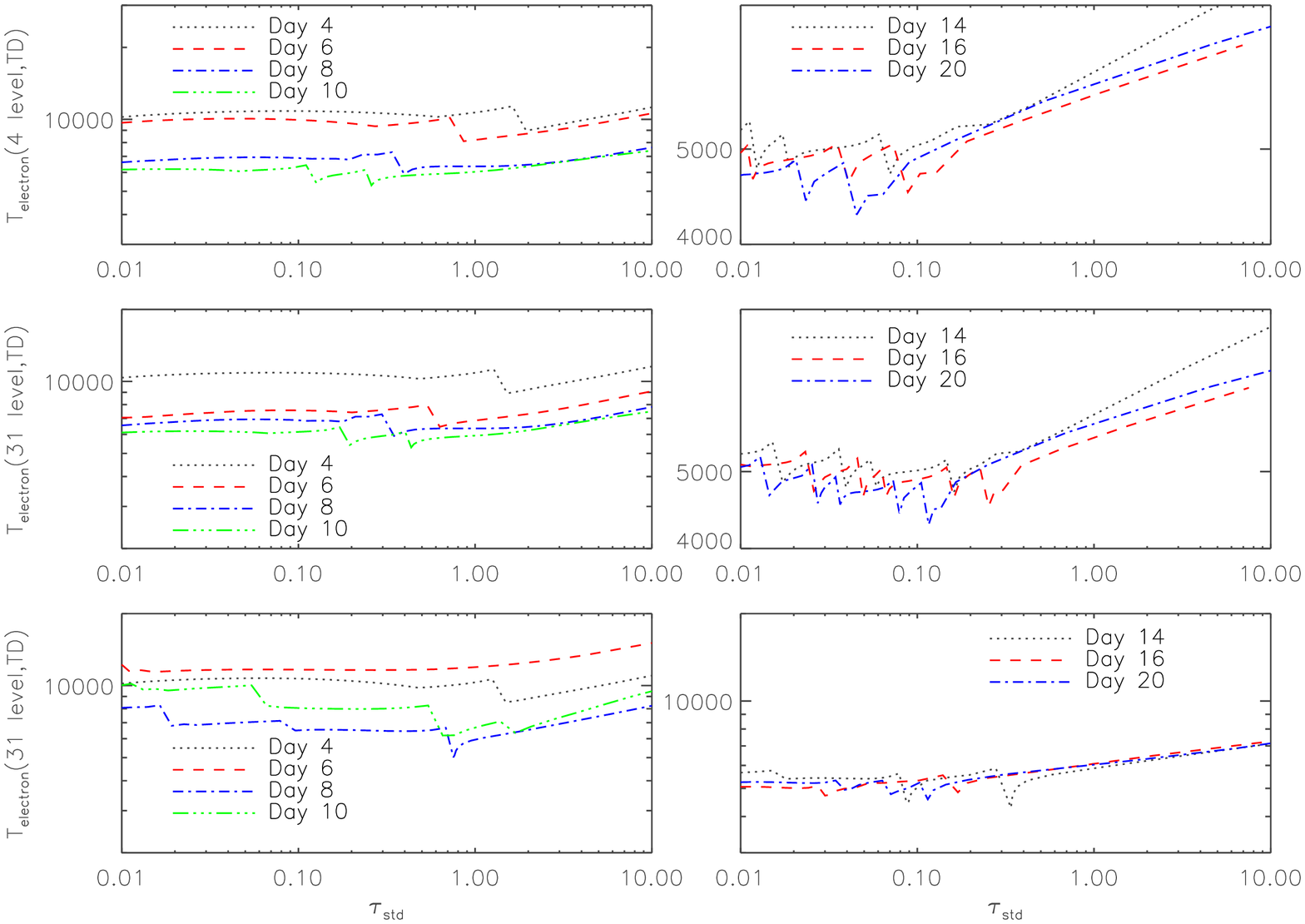}
\caption{Comparison between electron temperatures at different
  systems. The upper panel has 4-level hydrogen atom case. The middle
  panel has the multilevel time dependent case. The bottom panel has
  the multilevel time independent case. 
\label{temp_struc}}
\end{figure*}

\begin{figure*}
\centering
\includegraphics[width=0.65\textwidth,angle=90]{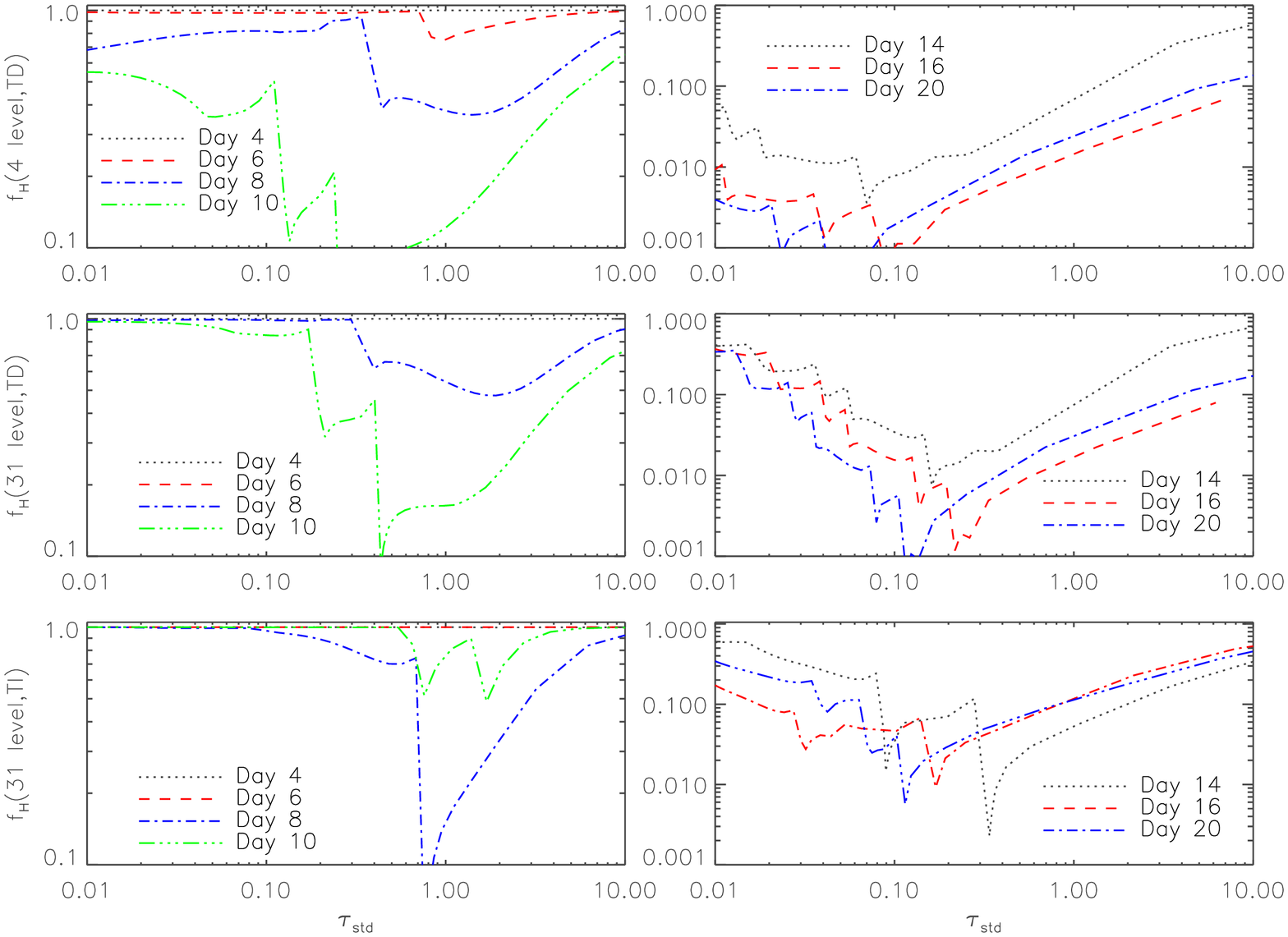}
\caption{Comparison between ionization fractions at different
  systems. The upper panel has 4-level hydrogen atom case. The middle
  panel has the multilevel time dependent case. The bottom panel has
  the multilevel time independent case. 
\label{ionization_frac}}
\end{figure*}

The upper panel shows the free electron density for Case 1 (where
only hydrogen is present with $n=2$ and rate equations are time
dependent) over different epochs.
Days 2--6 show a  steady rise in the free electron
density in the system as the optical depth becomes higher. The
electron density declines at later epochs due to cooling and geometric dilution.
These trends of high electron density at very early epochs
and lower electron density at the later epochs are 
expected from the fact that the 
temperature is high at early times.
At later epochs, the
electron density at a given optical depth falls off due to the
expansion and declining temperature. At epochs 4--6~d, the hydrogen in the system is
almost completely ionized and the free electron density is the highest
at this epoch. In the right  plot of the top panel of
Figure~\ref{electron_density}  we display the 
electron density for the epochs 14--20~d. At these epochs, hydrogen
ionization is lower and recombination of hydrogen begins. This
is evident from the corresponding temperature structure.
Figure~$\ref{temp_struc}$  displays the electron
temperature of the system, which controls the free electron density and
the level of ionization. The top left panel displays the
temperatures for early epochs ($T \sim 8000-10000$~K for epochs $<
8$~d). In the top right panel  the 
temperature drops to around 5000~K and this is where the transition
from the fully ionized to the partially ionized regime occurs. The
recombination regime begins roughly at day 8. Between optical
depths  $\tau_{\mathrm{std}} = 0.01-0.1$, the
temperature is around 
5000 K for epochs later than about 8~d since explosion,   thus
recombination starts to take 
place, reducing the free electron density. For  optical depths higher than
$\tau_{\mathrm{std}} =0.1$, the electron temperature is still high
enough (even at later epochs) 
to maintain a high level of ionization. Using the values of the free
electron density 
and the temperature structure it is easy to understand the ionization
fraction for 
hydrogen in the system which is displayed in 
Figure~$\ref{ionization_frac}$. 
The optical depth is indicated using
  the value $\tau_{\mathrm{std}}$ which is the optical depth in the
  continuum at  
5000~\AA.
Beginning at 6~d for 
Case 1, recombination starts to take place  at optical
depths $\taustd = 0.01-0.1$ causing a bumpy profile for the ionization
fraction.

For Case 2,  (time-dependent rate equations and
multi-level hydrogen with 31 bound levels) results
are displayed in the middle panel of our figures. We again refer
to the middle panel of the Figures
\ref{electron_density}--\ref{ionization_frac}. The basic nature 
of the temperature structure, free electron density, and the ionization
fraction are very similar to
Case 1. The main difference in this case is that for the later epochs,
the electron temperature is around 5000 K for optical depths $\taustd=
0.01-0.3$. Thus, the recombination front moves deeper into the expanding
ejecta
for the multilevel case with metals. This may be a 
cumulative effect due to the metals and the multiple bound levels. The
metals tend to suppress the ionization of hydrogen at higher
optical depths, hence we would expect the ionization front for hydrogen
to move deeper into the object. The multiple
bound levels enhance 
the recombination at any optical depth. The ionization fraction for
hydrogen follows the free electron density and the temperature profile
by producing a bumpy profile between for $\taustd=0.01-0.3$.

For Case 3, where the rate equations are time-independent with
the 31-level hydrogen model atom,  results are shown in 
the bottom panel of our plots. In this case the recombination
front moves deeper inside compared to Cases 1 and 2. Thus, the
electron temperatures of about 5000 K are reached at optical depths
$\taustd > 1.0$ for later epochs.

To summarize our results on the ionization fraction:
1) In all three cases we
observed that the results from the free electron density, electron
temperature and ionization fraction are consistent with each other.
2) As the supernova expands, the recombination front moves deeper
inside. The span of the recombination front is different in all three
cases, being largest in Case 3. Overall, the free electron densities
are similar in all three cases.  3) In the recombining regime, the
electron density drops, causing a drop in the ionization fraction as
well. 
  4) All three cases (31 level time dependent, 31 level time
  independent and 4 level time dependent) have visibly
  different ionization fraction profiles, but the difference decreases
  between Cases 2 and 3 (31 level hydrogen model). In other words 
  Case 1 has the ionization profile that is different from Case 2
  and Case 3. The 
  ionization fraction in these models after  10~days is higher than that
  in the 4 level  hydrogen atom time dependent model (Case 1).
5)In the recombination regime, there are glitches at the
  recombination front which probably arise from the imperfect spatial
  resolution of the ionization front. It is clear from the figures
  that the glitches do not affect the overall qualitative results.

\subsection{Spectral comparison}

Spectra are important as they are the only observable. In our
discussion we emphasize the $H_\alpha$ profile of the spectra. In
Figure~\ref{lineprofile_earlydays} we present the comparison between
the $H_\alpha$ line profile of the spectra for the Cases 2 and 3.  In
this plot we have days 4, 6, and 8.  We used \emph{dotted} lines for
Case 3 and solid lines for Case 2 in all our line profile and spectral
comparison plots.  The very early $H_\alpha$ profiles do not show any
differences between Cases 2 and 3. From Day 6 onwards we notice that
the line profile gets wider for Case 2 --- the time dependent case.
Figure~\ref{spectra_earlydays} shows a spectral comparison between
Cases 2 and 3. Overall, the spectral features for the Case 2 are
somewhat broader (especially between $4000-6500$~\AA).

We plot the $H_\alpha$ profile for  days 10, 14, 16, and 20 for 
Cases 2 and 3 in Figure~\ref{lineprofile_laterdays}.  We see a
similar broadening in the profile for Case 2 except at day 16.
This broadening becomes more obvious if one looks at the spectral
plot (Figure~\ref{spectra_laterdays}).  
The spectra seem to have somewhat broader
features for  Case 2, especially in the regime $\lambda <4500$~\AA\ and
$\lambda >8000$~\AA.
We emphasize that 
  these spectral profiles for SN 1987A were generated by independently
  tuning their luminosity to obtain the best fit to the observed
  spectra. To summarize we find that the spectral 
features become somewhat broader with time dependent treatment of the
rate equations, but the effect is smaller than that found by
\citet{dessart08}. 

Figure~\ref{spectra_observed} shows a comparison of the model spectra
with observations for SN~1987A at days 4 and 6. There is a noticeable
improvement in the $H_\alpha$ profile for the time dependent case
versus the time-independent case, although some of the effect is due
to variations in the luminosity which is a parameter in our
models. Thus, time dependent effects can be important for the early
epoch $H_\alpha$ profiles.

\subsection{Results on the recombination time}

\begin{figure*}
\centering
\includegraphics[width=0.65\textwidth,angle=90]{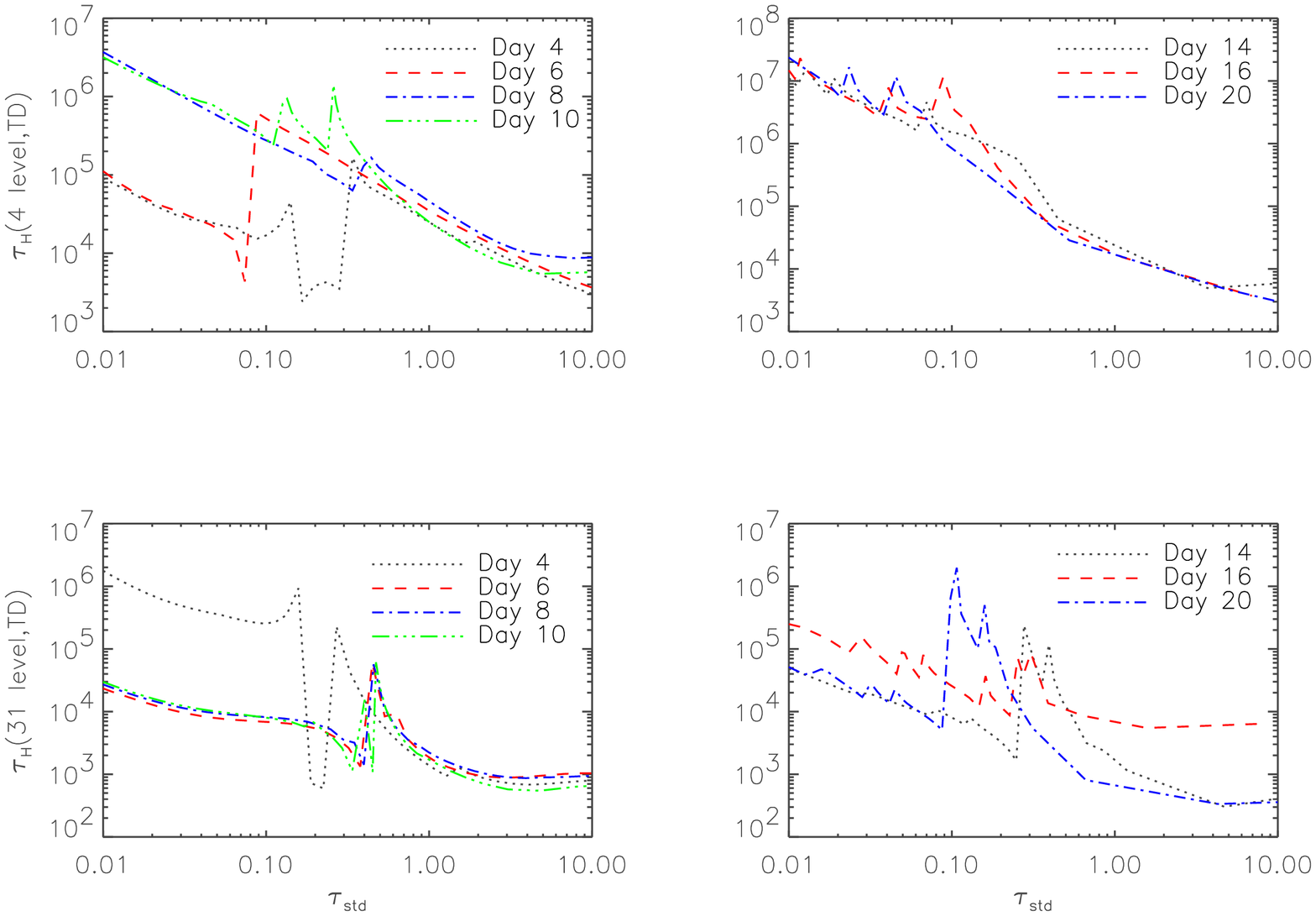}
\caption{Comparison between recombination times at different
  systems. The upper panel has 4-level hydrogen atom case. The middle
  panel has the multilevel time dependent case. The bottom panel has
  the multilevel time independent case. 
\label{recombination_time}}
\end{figure*}

\begin{figure*}
\centering
\includegraphics[width=0.65\textwidth,angle=90]{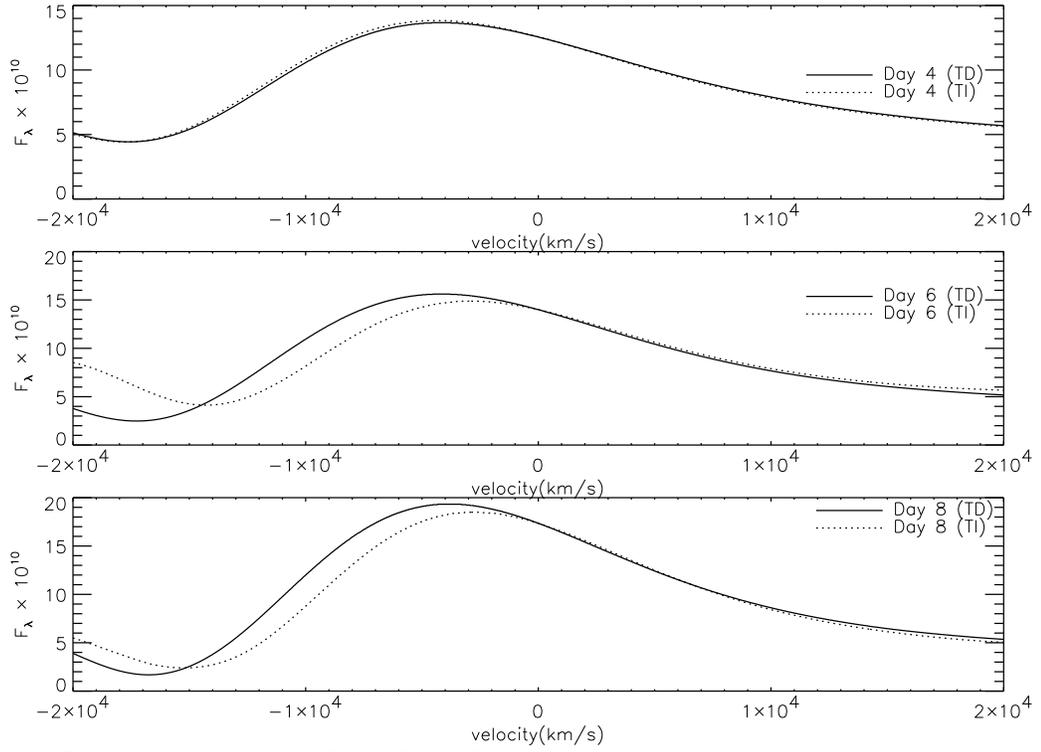}
\caption{Comparison of the line profiles of $H_\alpha$ for Days 4, 6, 8
  in the time dependent and time independent
  cases for a model that is appropriate to SN
    1987A. The luminosity in each case has been tuned to fit the observations.
\label{lineprofile_earlydays}}
\end{figure*}

\begin{figure*}
\centering
\includegraphics[width=0.65\textwidth,angle=90]{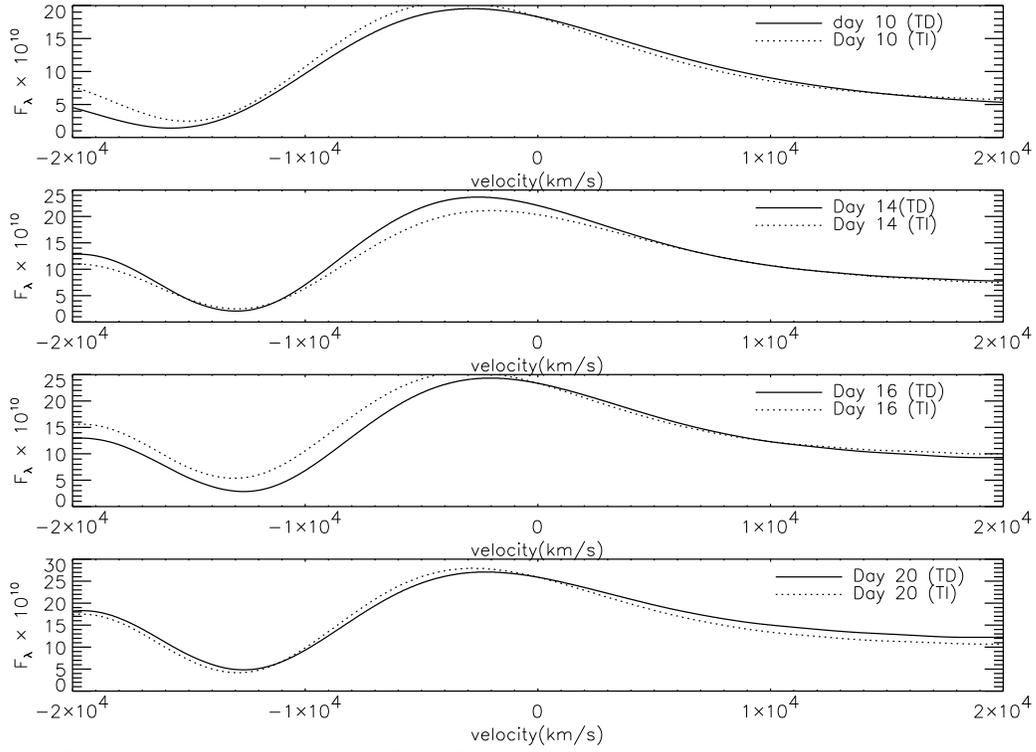}
\caption{Comparison of the line profile of $H_\alpha$ for Days 10, 14,
  16, 20 in the time dependent and time independent cases
  for a model that is appropriate to SN 1987A. The
    luminosity has been tuned to fit the observations in each case.
\label{lineprofile_laterdays}}
\end{figure*}

\begin{figure*}
\centering
\includegraphics[width=0.65\textwidth,angle=90]{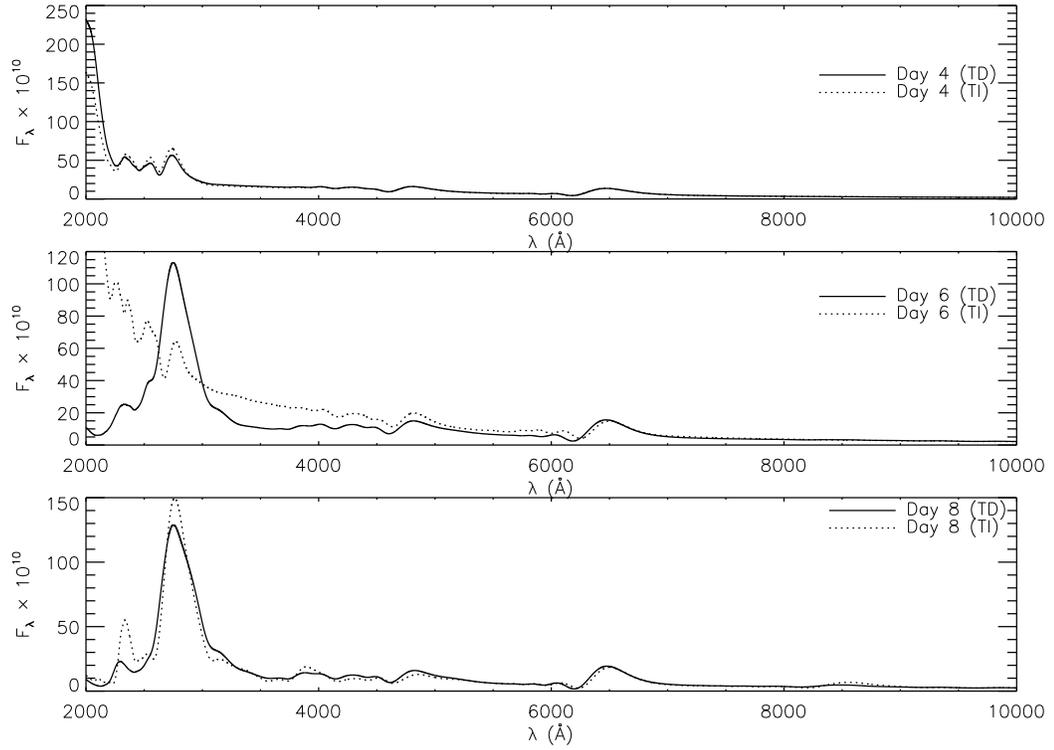}
\caption{Comparison of the spectra for Days 4, 6, 8 
  in the time dependent and time independent cases for a model that is
  appropriate to SN 
    1987A. The luminosity in each case has been tuned to fit the
    observations. The differences in the day 6 UV spectra are
    primarily due to
    variations in the opacity due to the exponential dependence
    of the Fe~II populations on the temperature and not to time
    dependence in the hydrogen rate equations.
\label{spectra_earlydays}}
\end{figure*}

\begin{figure*}
\centering
\includegraphics[width=0.65\textwidth,angle=90]{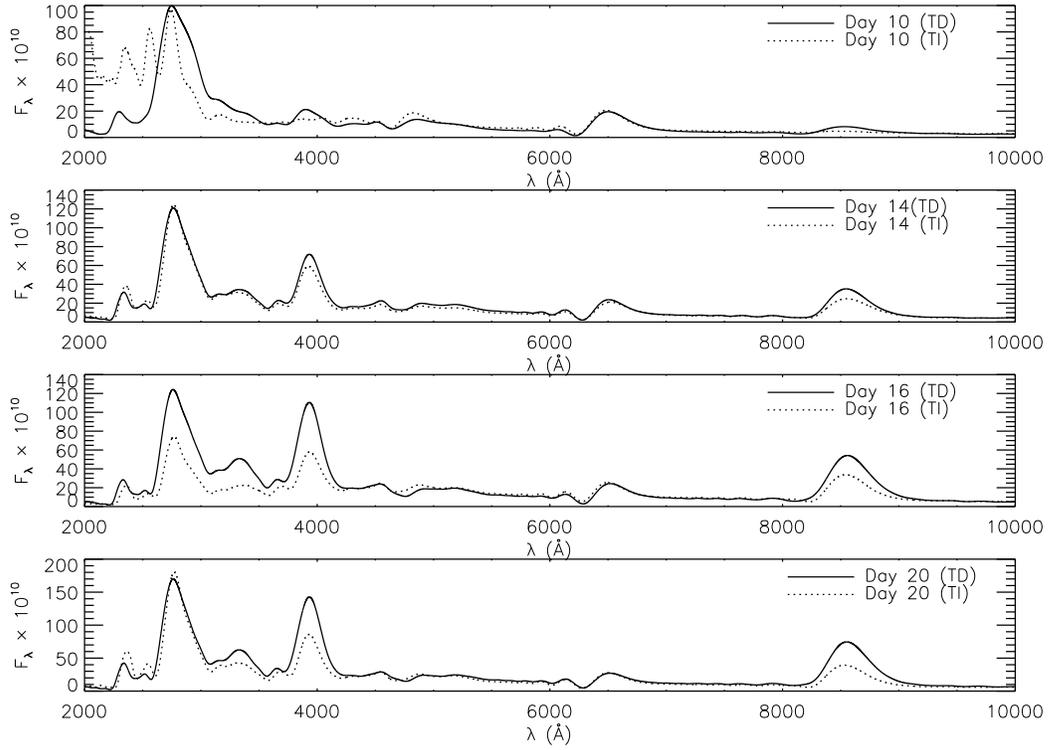}
\caption{Comparison of the spectra for Days 10, 14, 16, 20 in the time
  dependent and time independent cases for a model that is appropriate
  to SN 1987A. The luminosity in each case has been tuned to fit the
  observations.
\label{spectra_laterdays}}
\end{figure*}

\begin{figure*}
\centering
\includegraphics[width=0.65\textwidth,angle=90]{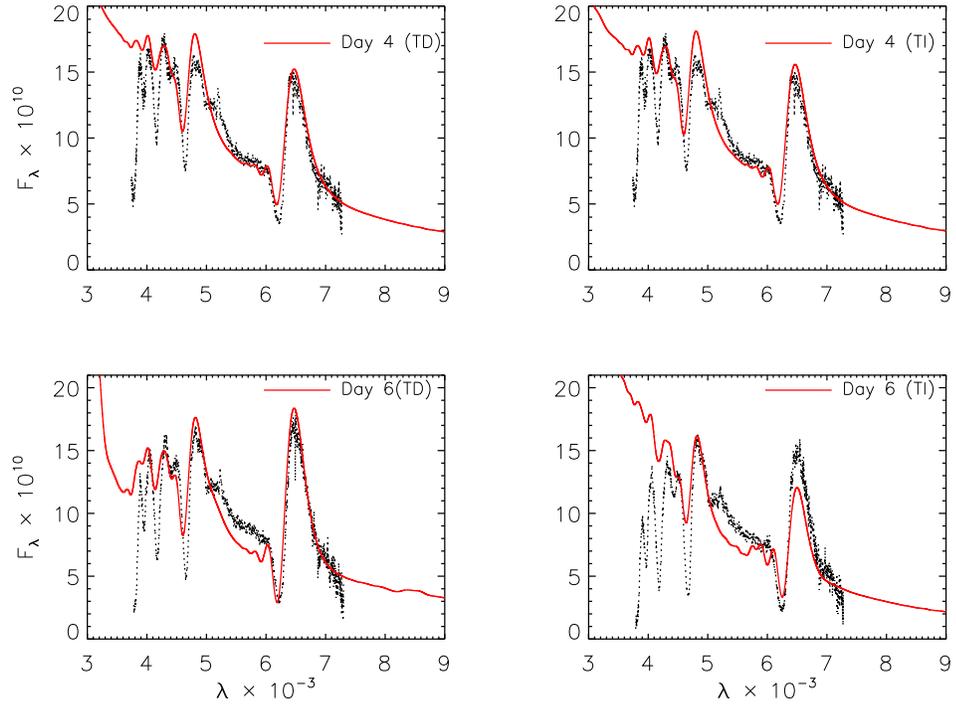}
\caption{Comparison of the spectra day 4 and 6 with time dependent and
  time independent cases with observed spectra of SN 1987A. The
  luminosity in each case has been tuned to fit the
  observations. The $H_\alpha$ profile is improved in the time
  dependent case.}
\label{spectra_observed}
\end{figure*}

The recombination time is a crucial quantity.  The motivation for this
paper was to study if the apparent importance of time dependence could
be verified from the numerical estimate of the effective recombination
time. It is important to check if the effective recombination time is
essentially comparable to the age of the supernova. Since our spectra
were generated by independently tuning the luminosity, it is important
to investigate the recombination time scale to be able to conclude if
the time dependence in the rate equations is indeed important.  As
noted above, the semi-analytic result is for a two-level atom and the
effects of both non-resonant de-excitations and a reduction in the
escape probability due to more channels could alter the numerical
values compared to the simpler analysis.
Figure~\ref{recombination_time} shows the recombination time for Cases
1 and 2.

Case 1, shown in the upper panel of Figure~\ref{recombination_time},
the recombination time 
(referred as $\tau_{H(4)}$) follows an inverted profile compared to
the ionization 
fraction (see Figure~\ref{ionization_frac}). This is
consistent with the fact that the low ionization fraction will reduce
recombination, resulting in a longer recombination time. Similar to
the profiles for the ionization fraction and the electron density, the
recombination time also gives a bumpy structure in the recombining
region. In Case 1, for epochs $>6$~d the recombination
time is around $10^{6}-10^{7}$ for $\taustd = 0.01-0.1$. Below the 
recombining regime, the recombination time is
lower. This is due to the increasing density.
At lower $\taustd$, the
recombination time increases because the
electron density declines.

The bottom panel of Figure \ref{recombination_time} shows the
recombination time for Case 2. The numerical value of the
recombination time, $\tau_{H(31)}$, is highest for the earliest
epochs.  This is because there is barely any recombination since the
temperature is $T > 10^{4}$ K (see Figure~\ref{temp_struc}).   After
the initial expansion, the electrons 
start to recombine. At the ionization front, due
  to the falling electron density, the recombination time increases at
  later epochs. Although outside of the ionization front the
  recombination time decreases with the expansion of the
  supernova. This is the case  both for 
  $\tau_{H(4)}$ and $\tau_{H(31)}$ except that this effect is more evident
  at later epochs in the multi-level case (bottom panel of Figure
  \ref{recombination_time}).  To summarize: 1) $\tau_{H(31)}$ and
$\tau_{H(4)}$ follow the profile expected from their respective
free electron density and temperature profiles and  almost follow an inverted
profile compared to the ionization fraction profile. In other words, the
recombination time is found to be proportional to the free electron
density and temperature and inversely proportional to the ionization
fraction.

2) The recombination time is higher for low optical depths. There
is a break in the monotonic decline in the recombination time (as it
moves toward higher optical depth) in the recombining
regime. There is a rise in the recombination time in this
regime. For optical depths $\taustd > 1.0$, both $\tau_{H(31)}$ and $\th$
decrease monotonically due to the increase in the free electron density.

3) As the supernova expands, the recombination time at a given optical
depth decreases at almost all optical depths (Figure \ref{recombination_time}).

4) Figure \ref{tanal_by_th31} 
  compares the  time-dependent multi-level hydrogen atom
  and the classical recombination time, $\tau_{anal}$.  In
  the optically thin regime ($\tau_{std} <0.10$)  $\tau_{H(31)}$
  is close to $\tau_{anal}$.
  Only right at the ionization front doe the ratio
  deviate significantly from unity and $\tau_{H(31)}$ is larger than
  $\tau_{anal}$. 

\section{SN 1999\lowercase{em}}
SN1987A is one of the best observed astronomical objects. It is also a
peculiar Type II supernova. It has two maxima in its light curve and
it also has very steep rise to its maximum.  On the other hand SN
1999em has steady rise to the maximum and a well-defined
plateau. Thus, we also studied the effects of time-dependence in
SN~1999em.  We began with the day 7 spectrum and marched forward in
time until around day 40 (since explosion). We again considered a 31
level hydrogen atom and treated only hydrogen in NLTE.  The departure
coefficients (referred as $b_{i}$) are the ratio of the real level
population density to the expected LTE level population density
\citep{MC37,mihalas78sa}. Figures~\ref{bi_99em_n1}--\ref{bi_99em_n3}
display the departure coefficients for the levels $n=1$ (ground
state), $n=2$, and $n=3$.  It is evident from our plots that time
dependence is important at early times although the effect is very
small.  This effect diminishes with time. Also the effect is only
relevant in the recombining regime. The explanation of this transient
phenomenon is easier to understand in the Lagrangian frame. In a given
mass element at very early times if the temperature is much higher
than 5000~K, the hydrogen is nearly completely ionized and the
$\frac{d}{dt}$ term is very small.  At very late times when the
temperature falls much below 5000~K, the hydrogen is mostly neutral
and the $\frac{d}{dt}$ term is also small.  Thus, the $\frac{d}{dt}$
term is important in the recombination regime.  The effect decreases
with time.  Our results agree with those of
\uc05. 
   The calculations are performed on a Lagrangian grid, however, the
   results are best presented on the $\tau_{std}$ grid.
  As an aid to  interpreting these figures, Figure
  \ref{v_vs_tau} shows the profile of velocity 
   vs $\tau_{std}$. We find that the change in the departure
  coefficients for the ground state is up to 12\%. For the first and
  second excited states the change in the departure coefficients is
  up to 4\% and 2\% respectively. The change was found to be maximum at
  day 10.  In order to test the sensitivity of our
results to the timestep we compared our results for SN~1999em for the
case where we use a 4 day interval to that where we use 1 day interval
in the time dependent rate equation for hydrogen. We find
in the 14~day spectra, the change in the departure
coefficients (derived from two different timesteps) is within 3\% (for the
ground state) in the line forming region. This effect becomes even
smaller at later times.

\section{Discussion and Conclusions}

We have studied the importance of 
time-dependence in the rate equations in a type II supernova
atmosphere due to the increase in the recombination time. 
We also
explored the effects of having many angular momentum sub-states.
In our description we use $\tau_{H(4)}$ and $\tau_{H(31)}$ as the recombination time for a 4-level hydrogen atom model and 31-level hydrogen atom model respectively.

1) \uc05 proposed that $\tau_{H(4)}$ is of the order of $10^{7}$~s at $T \sim
5000$~K and $n_{e}=10^{8}$ cm$^{-3}$. For epochs later than 6 days, we
calculated $\tau_{H(4)}$ 
at similar electron density and temperature to be around $10^{7}$~s
($\taustd \sim 0.02$).  Thus, we 
recover the
recombination time 
scale predicted by \uc05 for the 4-level hydrogen case. 
An important difference between our results and those of \uc05 is
that  we find the 
electron densities and temperatures which they used as photospheric
conditions to occur
at $\taustd \sim 0.02$ and not near $\taustd = 1.0$. Around
$\taustd=1.0$, the free electron density even for Case 1 is much 
higher and hence the recombination time is much smaller. This
optical depth mismatch is due to the fact that we are looking at the 
continuum optical depth and not the line optical depth.
The ratio of the Balmer line opacity to the
continuum opacity  is $\sim 10^{3}$ in the recombining
region.

We also compare the classical 
recombination time (calculated using
$1/\alpha n_{e}$)  with our calculated $\tau_{H(4)}$ generated
from the solution of the radiative transfer equation. We refer to the
approximate classical recombination time 
as $\tau_{anal}$. This allows us to determine 
the factor $w_{21}$ (\uc05).
Figure~\ref{tanal_by_th} displays the ratio
$\tau_{anal}/\tau_{H(4)}$,  a direct measure of $w_{21}$ described by
\uc05.

\begin{figure*}
\centering
\includegraphics[width=0.65\textwidth,angle=90]{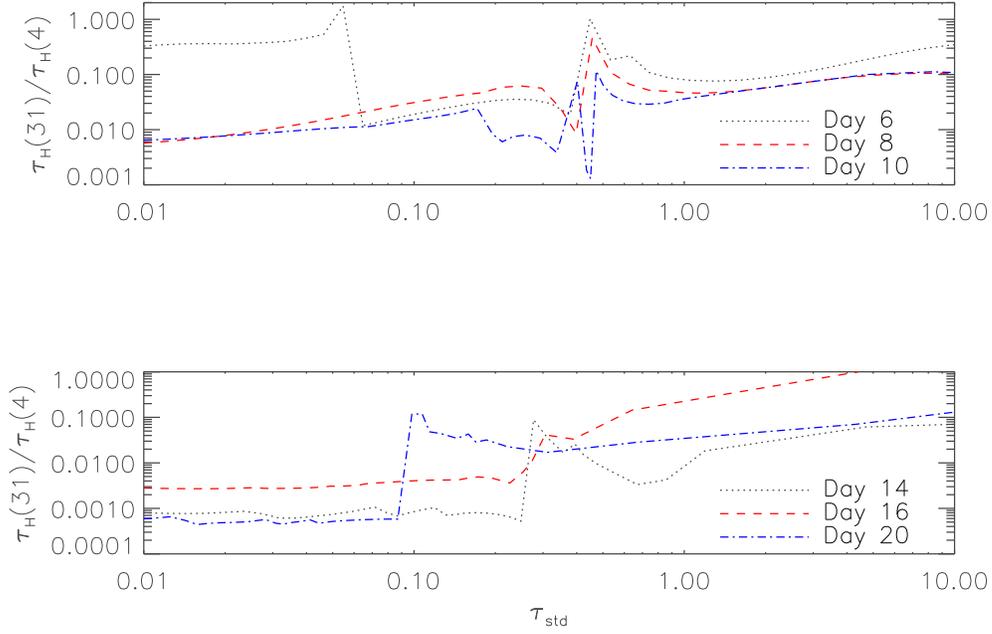}
\caption{Ratio of recombination times for the time dependent
  multilevel hydrogen atom case and the 4-level hydrogen
  atom model calculated at different days. The upper panel shows the
  earlier epochs and the lower panel the later epochs.
\label{ratio_recombination_time}}
\end{figure*}

\begin{figure*}
\centering
\includegraphics[width=0.65\textwidth,angle=90]{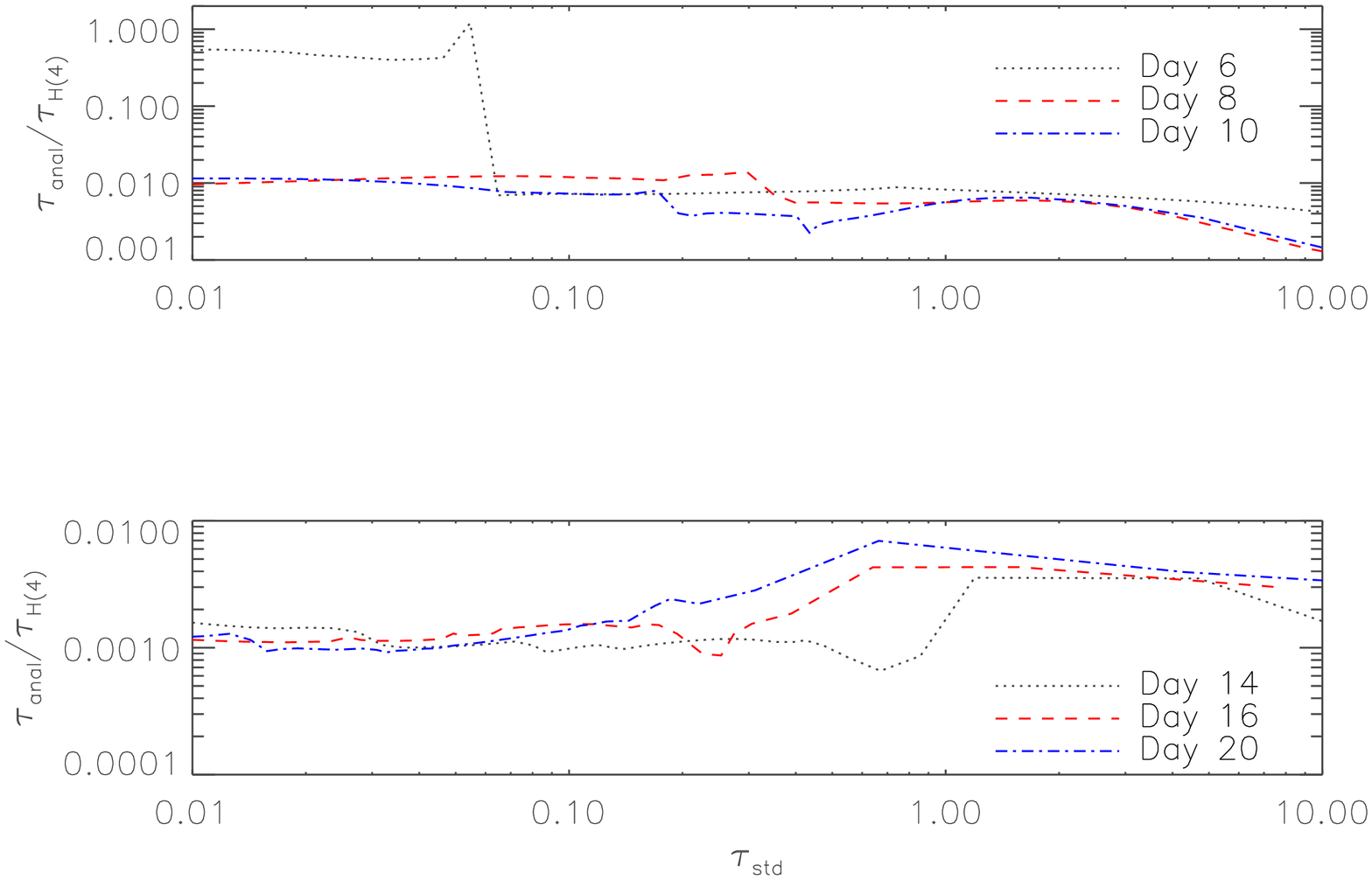}
\caption{Ratio of recombination times for the analytical
  calculation and the 4-level hydrogen atom model calculated at
  different epochs. The upper panel shows the
  earlier epochs and the lower panel the later epochs.
\label{tanal_by_th}}
\end{figure*}

\begin{figure*}
\centering
\includegraphics[width=0.65\textwidth,angle=90]{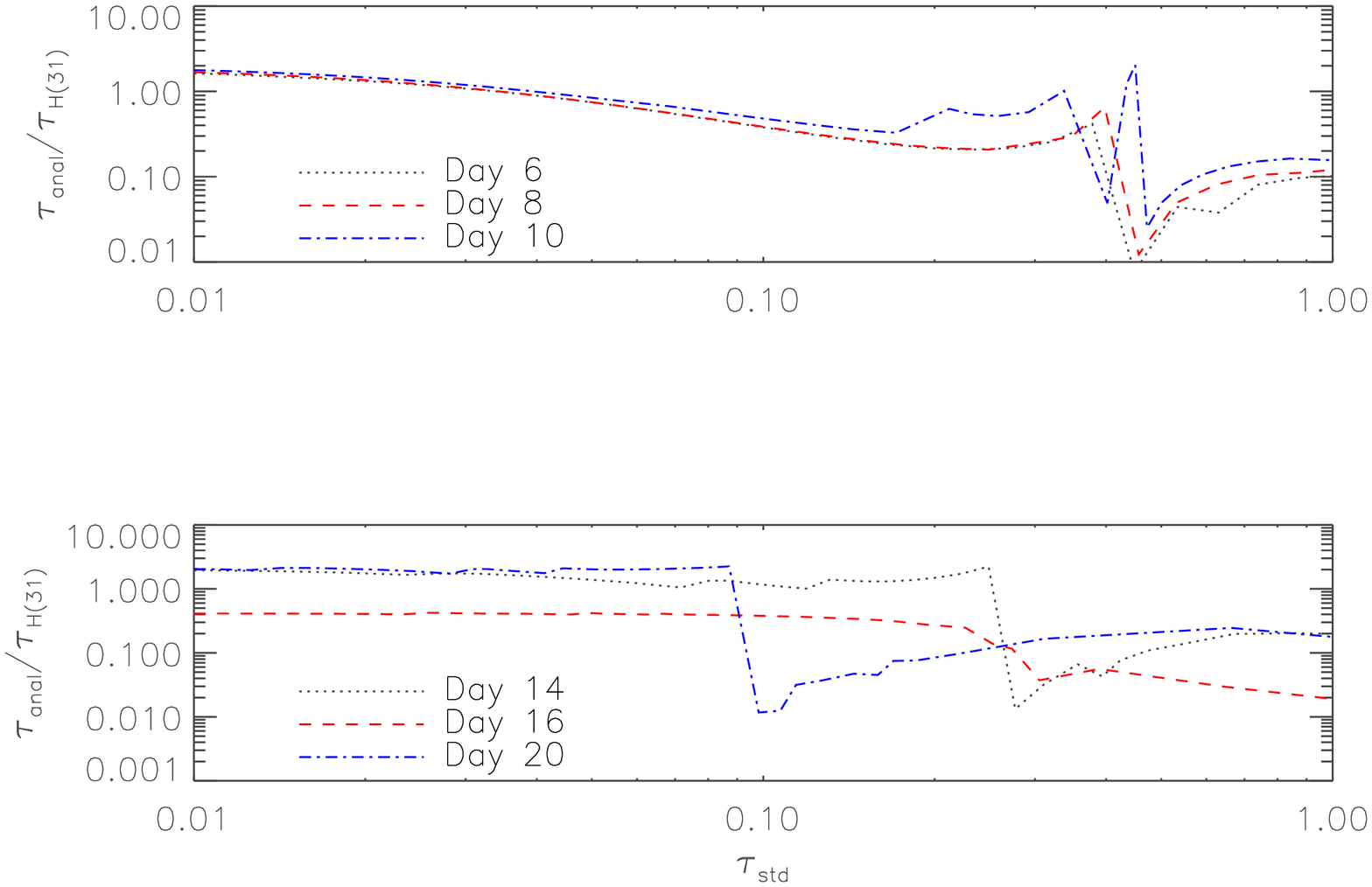}
\caption{Ratio of recombination times for the analytical
  calculation and the 31-level hydrogen atom model calculated at
  different epochs. The upper panel shows the
  earlier epochs and the lower panel the later epochs.
\label{tanal_by_th31}}
\end{figure*}

\begin{figure*}
\centering
\includegraphics[width=0.65\textwidth,angle=90]{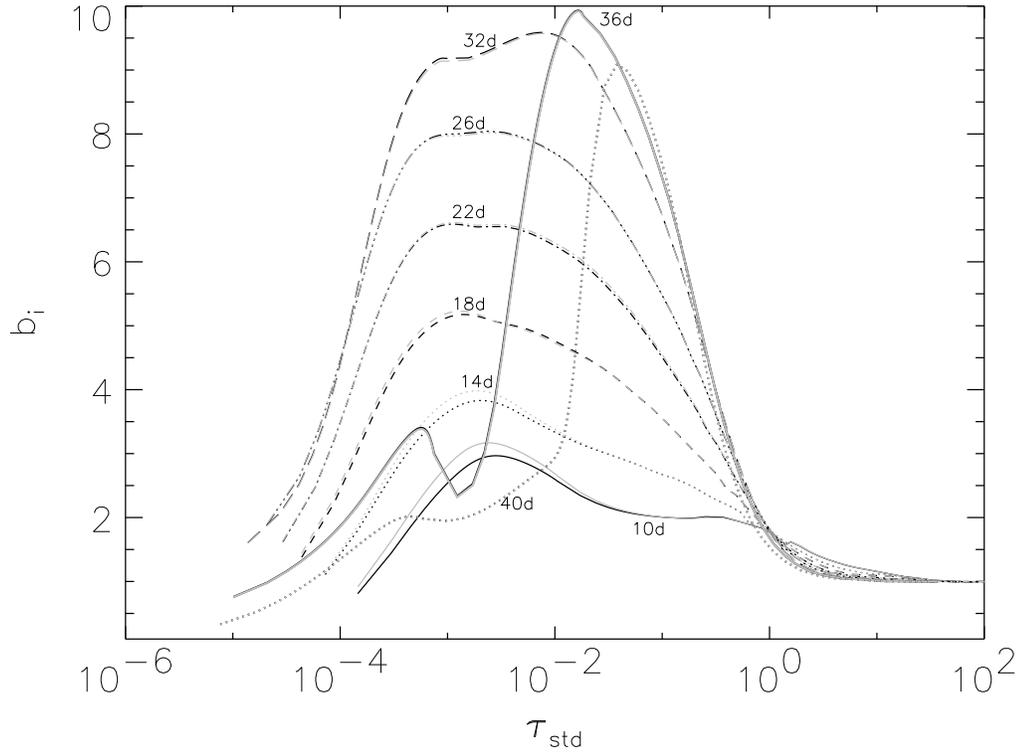}
\vspace*{20pt}
\caption{Departure coefficient, $b_{1}$, as a function of $\tau_{std}$ for the ground
  state of hydrogen. The black line is 
for the time dependent case and grey line is the time independent
case for each day. 
\label{bi_99em_n1}}
\end{figure*}

\begin{figure*}
\centering
\includegraphics[width=0.65\textwidth,angle=90]{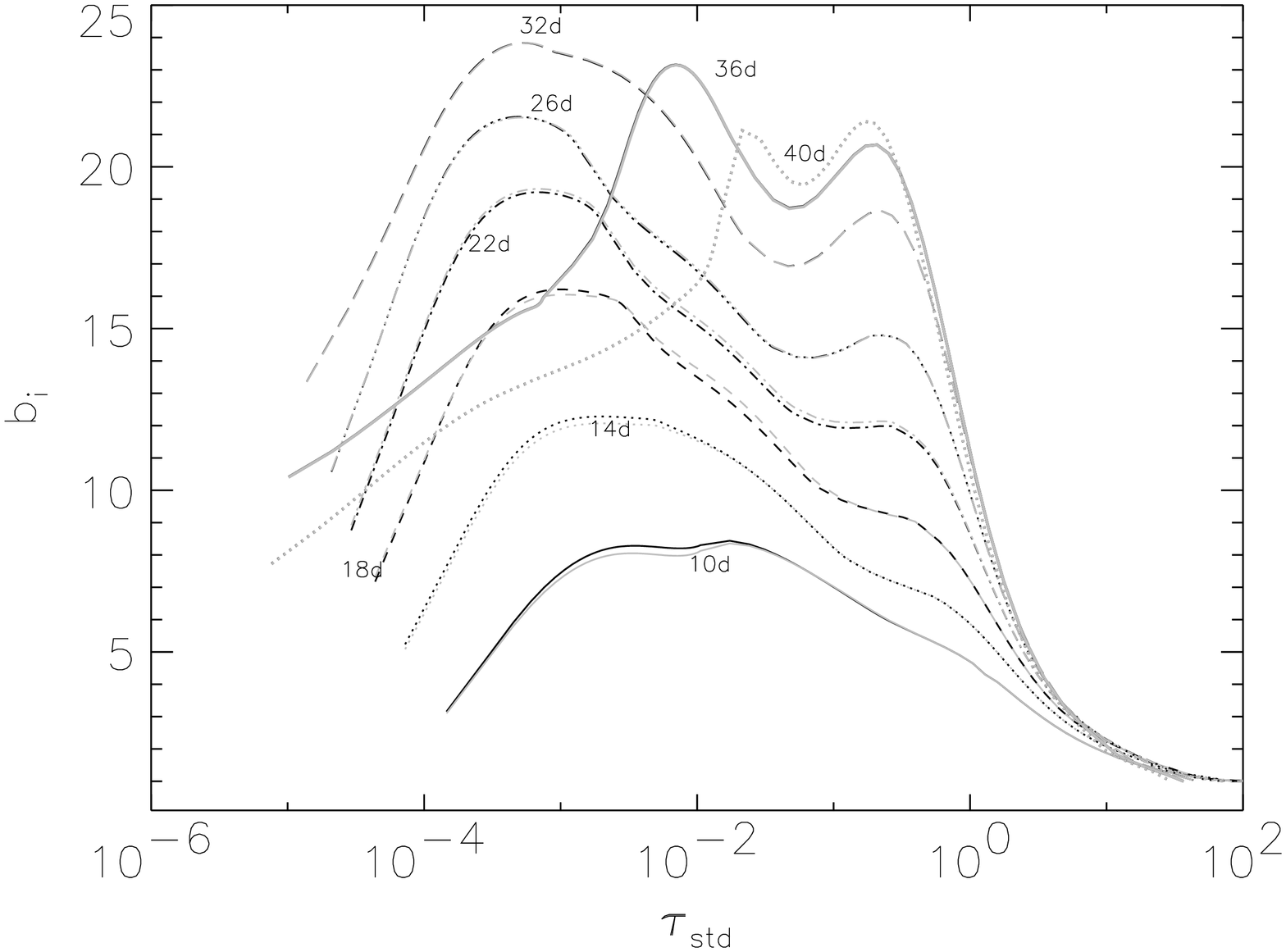}
\vspace*{20pt}
\caption{Departure coefficient, $b_{2}$, as a function of $\tau_{std}$ for the $n=2$ state of
  hydrogen. The colors and line-styles are the same as  
in Figure \ref{bi_99em_n1}.
\label{bi_99em_n2}}
\end{figure*}

\begin{figure*}
\centering
\includegraphics[width=0.65\textwidth,angle=90]{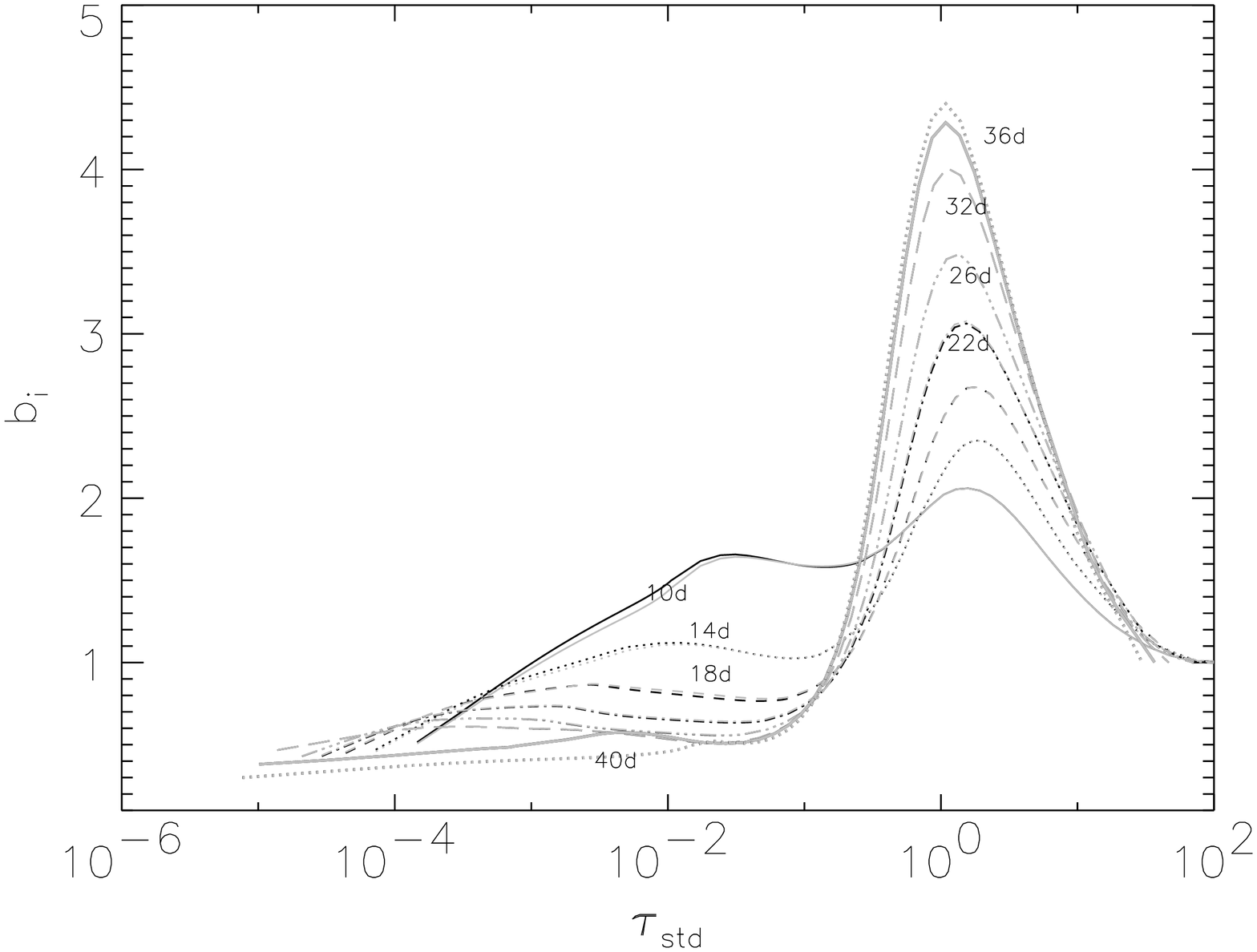}
\vspace{30pt}
\caption{Departure coefficient, $b_{3}$, as a function of $\tau_{std}$
  for the $n=3$ state of hydrogen. 
The colors and line-styles are the same as  
in Figure \ref{bi_99em_n1}.
\label{bi_99em_n3}}
\end{figure*}

\begin{figure*}
\centering
\includegraphics[width=0.65\textwidth,angle=90]{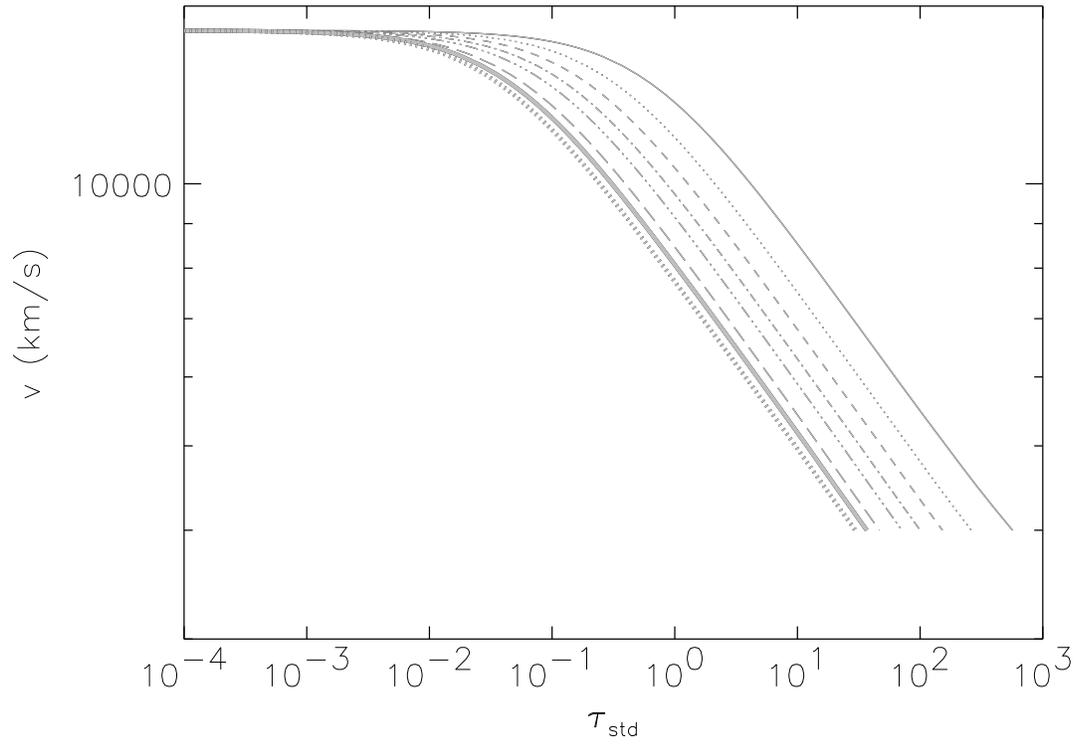}
\vspace{30pt}
\caption{Expansion velocity of each layer as a function of $\tau_{std}$
  for the SN 1999em model. 
The colors and line-styles are the same as  
in Figure \ref{bi_99em_n1}.
\label{v_vs_tau}}
\end{figure*}

In Figure~\ref{ratio_recombination_time}, we display the
ratio of the recombination time for Case 2 to the
recombination time from a simple  hydrogen  atom (Case 1)  at the same
epoch and at the same optical depth. 

1) For all epochs, $\tau_{H(31)}$ is smaller than $\tau_{H(4)}$ at
almost all optical depths. For higher optical depths, the ratio
increases and approaches unity.  because the ionization fraction for
hydrogen and $\tau_{H(31)}$ becomes independent of the number of bound
levels.  At this point the system is almost in LTE and the number of
free electrons depends only on the temperature.  2)
  Figure \ref{tanal_by_th} shows that the simple hydrogen atom model
  (4-level) overestimates the recombination time by a factor of 100 or
  more at all almost all epochs and optical depths.  3) Figure
  \ref{tanal_by_th31} shows that the recombination time recovered from
  a multi-level hydrogen atom compares with the standard value
  ($\tau_{anal}$) for $\tau_{std}<1.0$. At the ionization front
  $\tau_{H(31)}$ differs significantly from $\tau_{anal}$.
  4) Based on our work on SN 1999em, we find that the transient nature
  of the level population of hydrogen is not a crucial factor on the
  plateau, but it is important in the active recombining regime. This
  effect could be important for early features in the spectra but the
  steady state approximation should be valid in the plateau
  phase.  5) In the early early spectra of SN~1987A we find that
  $H_\alpha$ is better fit using the time-dependent formulation.  6)
  The recombination time jumps above the underlying 
   trend only at the ionization front.
  This is just caused by the sudden reduction of the electron
  density driving the the recombination time
  higher in the ionization front.

It is very difficult to decouple the effects of time
   dependent rate 
  equations in a radiative transfer framework and the effects of
  multi-level model atoms
  by only looking at
  spectra. The Balmer profile of the spectra of SN 1987A at early times
  (Figure \ref{spectra_observed}) and the level populations from our
  results on SN 1999em (Figure \ref{bi_99em_n1} -- \ref{bi_99em_n3}),
  indicate the effects of the time dependent rate equations at early
  times. 
  The increased recombination time (Figure
  \ref{recombination_time}) at early times for SN 1987A also supports
  the importance of time-dependent rate equations.
 Therefore we conclude 
  that time dependence is more important at early
  times than later times. The effect of
  multi-level atoms can be seen in
  Figure \ref{recombination_time} which clearly shows than even at
  later times $\tau_{H(4)}$ is much higher than $\tau_{H(31)}$.
  Therefore conclusions using only a 
  4-level model may overestimate the importance of time dependence.
  Of course, including time-dependence
  is not terribly costly numerically and thus it can be included when
  a time sequence is calculated.

\section*{Acknowledgments} 
This work was supported in part by NSF grant AST-0707704, Department
of Energy Award Number DE-FG02-07ER41517, and by SFB grant 676 from
the DFG.  This research used resources of the National Energy Research
Scientific Computing Center (NERSC), which is supported by the Office
of Science of the U.S.  Department of Energy under Contract No.
DE-AC02-05CH11231; and the H\"ochstleistungs Rechenzentrum Nord
(HLRN).  We thank both these institutions for a generous allocation of
computer time.

\bibliography{apj-jour,refs,bib_recom,baron,sn1bc,sn1a,sn87a,snii,stars,rte,cosmology,gals,agn,atomdata,crossrefs}

\clearpage

\end{document}